\documentclass{article}
\usepackage{emulateapj}
\usepackage{float,epsfig,psfig}
\usepackage{pstricks}
\def\hea4{{\it HEAO~A4}}
\def\heaoa2{{\it HEAO~A2}}
\def\heao1{{\it HEAO~1}}

\def\h0{$H_{\rm o}=50$~km~s$^{-1}$~Mpc$^{-1}$}
\def\q0{$q_{\rm o}$}

\def\cms3  {~{cm$^{-3}$}}

\newcommand{\mincir}{\raise
  -2.truept\hbox{\rlap{\hbox{$\sim$}}\raise5.truept \hbox{$<$}\ }}
\newcommand{\magcir}{\raise
  -2.truept\hbox{\rlap{\hbox{$\sim$}}\raise5.truept \hbox{$>$}\ }}

\begin{document}

\submitted{ApJ, 2004, Aug. 20 issue}
\title{XMM-Newton study of A3562 and its immediate Shapley environs.}
\author{{A. Finoguenov$^{1,2}$, M. J. Henriksen$^2$, U. G. Briel$^1$,
J. de Plaa$^3$, J.S. Kaastra$^3$}} 
\affil{
{$^1$ Max-Planck-Institut f\"ur extraterrestrische Physik,
             Giessenbachstra\ss e, 85748 Garching, Germany}\\
{$^2$ Joint Center for Astrophysics, Physics Department, University of
  Maryland, Baltimore County, Baltimore, MD 21250, USA}\\ 
{$^3$ SRON, National Institute for Space Research, Sorbonnelaan 2, 3584 CA
  Utrecht, Netherlands} }

\authoremail{alexis@jca.umbc.edu}

\begin{abstract}

Using the mosaic of six XMM-Newton observations, we study the hydrodynamic
state of the A3562, a cluster in the center of the Shapley Supercluster.
The X-ray image reveals a sharp surface brightness gradient within the core
of A3562, a 200 kpc ridge extending to the south-west. A nearby group,
SC1329-313, imaged within this program also exhibits a tail aligned with the
ridge. Study of the pressure and entropy identified the ridge with a 40\%
pressure enhancement. An associated Mach number of 1.15 relative to a
polytropic sound speed for a 5 keV plasma requires the velocity of the group
to be 1400 km/s, while the projected velocity difference between the cluster
and the group amounts to 1200--1500 km/s. Apparent disruption of the group,
if attributed to the ram pressure, also requires the velocity of the group
in the cluster frame to amount to $1700\pm150$ km/s. The sharp surface
brightness gradient at the center is identified with a contact
discontinuity, which together with dove tails in the the large-scale entropy
distribution reveals a sloshing of the BCG in response to the passage of the
SC1329-313 group. Using the extent of the low-entropy tails in A3562 we
estimate the amplitude of sloshing motion to be $200h_{70}^{-1}$ kpc and the
oscillation frequency 1 Gyr. 

\end{abstract}

\keywords{galaxies: intra-galactic medium; clusters: cosmology; cosmic
  star-formation} 

\section{Introduction}

Superclusters or second-order clusters of galaxies as they were first dubbed
by George Abell (1958) are the largest coherent objects in the
Universe. They are only 5 to 50 times denser than the field so that
superclusters are not relaxed and should thus bear the imprints of the
physical processes that are dominant during their formation. Their mildly or
modestly nonlinear character offers direct clues about the ongoing growth of
cosmic structures in the Universe. In the past few years observational
evidence has increased in favor of a hierarchical formation of structure
also on supercluster scales (Rines et al. 2001). From optical and X-ray data
it was found that the core region of the Shapley Supercluster is close to
turnaround, which is a rather unique state for a supercluster, as only a
small fraction of them is expected to have collapsed by now.  A mosaic of
the A3562 cluster and its immediate environment was proposed for XMM-Newton
to search for the cluster interaction within the supercluster, to understand
its role in cluster formation within the supercluster. This also has an
important link to studies of the origin of the non-thermal emission.

As will become clear in the course of our investigation, A3562 exhibits core
oscillations in response to the passage of the SC1329-313 group, observed in
the current XMM-Newton program. So, this paper is devoted to a detailed
description of the disturbance the core oscillation brings to the
cluster. We identify pressure features and provide their interpretation in
connection to the interaction scenario for A3562. Our data on the SC1329-313
group show an interestingly high entropy level, suggesting that groups in
superclusters are systematically underluminous. We attribute it to a high
entropy of the intracluster gas in superclusters, possibly connected to the
structure formation shocks. Cosmic shocks emerge during structure formation
in the universe, due to gravitationally driven supersonic gas infall onto
collapsing objects. We also note a high elemental abundance in the outskirts
of both A3562 and the SC1329-313 group, which indicate an enhanced
importance of feedback effects. This has important consequences for studies
of the luminosity function of objects in superclusters.

We adopted the $D_L=215$ Mpc, $D_A=195$ Mpc and 57 kpc/arcminute
plate-scale, which for $z=0.0483$ of the cluster corresponds to
$H_0=70$~km~s$^{-1}$~Mpc$^{-1}$, and $\Omega_M=1-\Omega_\Lambda=0.3$.

\subsection{Overview of the previous X-ray results on A3562}

A3562, as a part of Shapley supercluster, received a wealth of observing
time at X-ray wavelengths. The Einstein observatory provided a first
temperature estimate (Breen et al. 1994), joint Einstein and ROSAT analysis
provided detailed mass estimates (Ettori et al 1997). A filamentary X-ray
structure on a several degree scale was reported by ROSAT (Kull \&
B\"ohringer 1999). Spectral analysis of ROSAT data, revealing a complex
temperature structure of A3562 has been presented in Bonamente et
al. (2002). A chain of clusters including A3562 and the SC1329-313 group has
been covered by both ASCA (Hanami et al. 1999) and BeppoSAX (Ettori et
al. 2000, Bardelli et al. 2002).

In view of the observational results and their interpretation, reported in
this paper, several results of the previous satellites are particularly
important, so we list them here. Of primary concern to this paper is the
temperature structure of A3562, which according to our XMM-Newton results,
is hotter to the south and colder to the north. This matches very well the
ASCA GIS results of Akimoto et al. (2003, see their Fig.8). The temperature
range, reported by ASCA for the central $11^{\prime}$ is $5.0\pm0.2$ keV
(Hanami et al. 1999) and by BeppoSAX LECS+MECS for the central $8^{\prime}$
is $5.1\pm0.2$, and a radial gradient in temperature from $\sim5$ keV in the
central 5 arcminutes to $\sim 3$ keV at radii exceeding 10 arcminutes
(Ettori et al. 2000). The ROSAT PSPC temperature for the central
$6^{\prime}$ is $3.7\pm0.4$ keV (Ettori et al. 2000).

Our XMM results reveal temperature variations in the 3.8--5.1 keV range
within the central $10^{\prime}$, so various temperature measurements
reported before are understood as due to a different sensitivity of the
detectors to underlying temperature structure. XMM results also show that
the temperature structure is asymmetric and thus is not removed in an
analysis using concentric annuli. The high sensitivity of XMM to the soft
band yields intermediate temperatures, typically $4.3\pm0.1$ keV, when
averaged by mass of the gas.

However, the residuals in the hard energy band, claimed in ASCA analysis,
could partially rise from the bright quasar, $14^{\prime}$ off center.

Prevalence of 3 keV gas outside the central $10^\prime$ of A3562, reported
by Ettori et al. (2000) is also seen in XMM data.

With respect to a nearby to A3562 group, SC1329-313, which is also a subject
of our XMM observation, the ASCA observation results in a temperature
estimate of $4.2\pm0.2$ keV within the central $11^{\prime}$ (Hanami et
al. 1999), while BeppoSAX yields $3.5\pm0.3$ keV within the central
$8^\prime$ (Bardelli et al. 2002). The XMM observation yields $3.1\pm0.1$
keV within the central $11^\prime$ (most of the counts come from eastern and
northern side of the group). As ASCA hardness ratio maps show (Akimoto et
al. 2003), this group exhibits substantial color variation, with the center
of the group being colder, while $8^\prime$ to the west there is a hot
spot. The XMM data do not cover the area of hot emission, and the BeppoSAX
extraction area only covers it partially.  This may explain the range in the
reported temperatures.

\begin{figure*}
\includegraphics[height=18.cm,angle=-90]{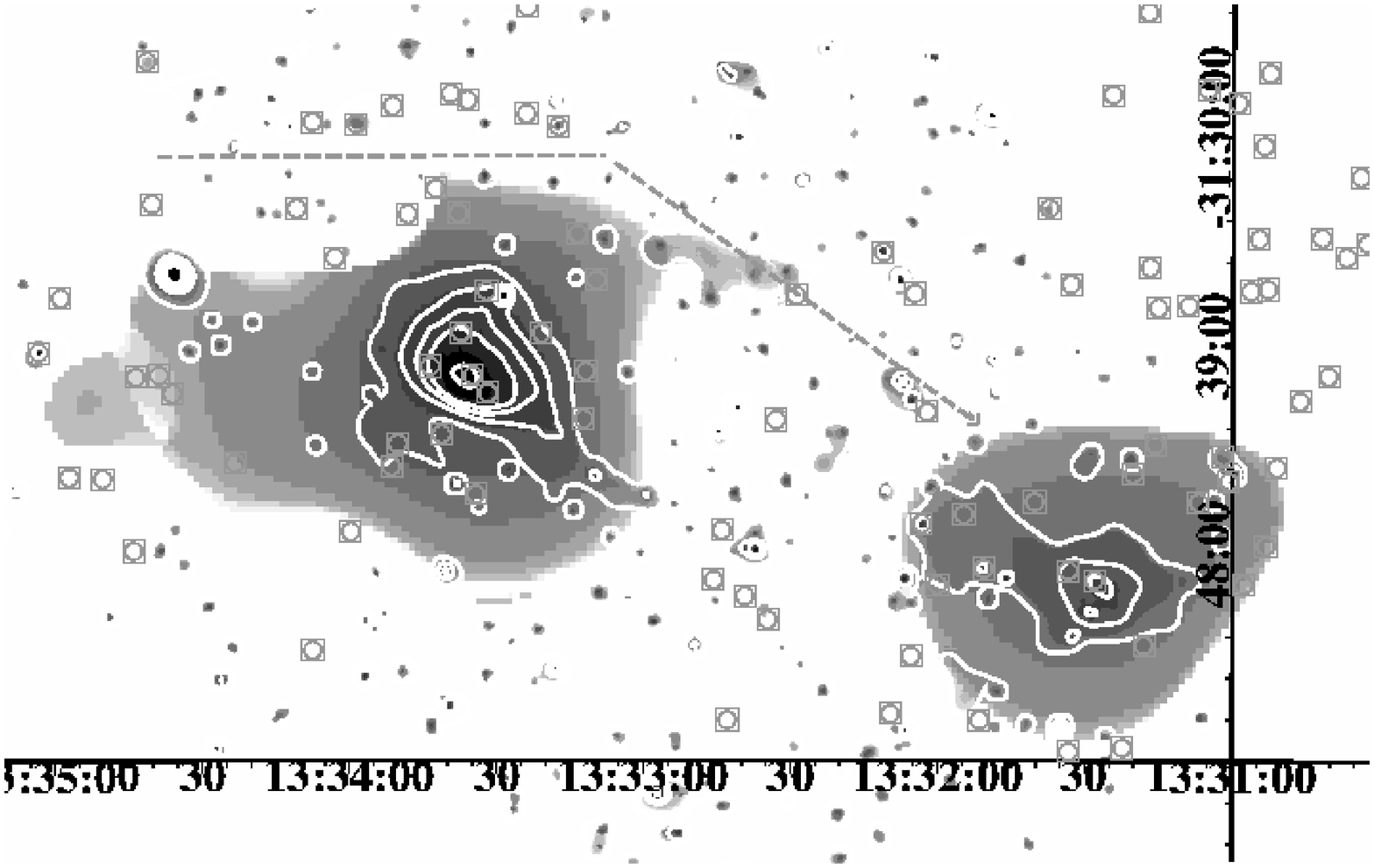}\\
\includegraphics[width=19.cm,angle=-180]{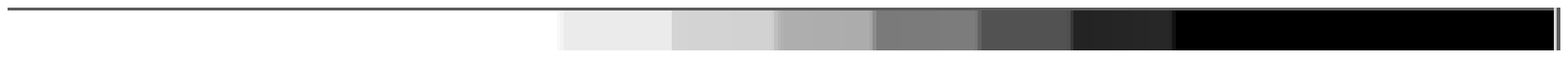}

\figcaption{Wavelet-reconstructed XMM-Newton image of A3562, obtained in the
0.8--2 keV band. The gray symbols indicate confirmed member galaxies. White
contours show the emission on small-scales. Gray dashed line, which ends
with an arrow, points to the direction of by-pass of the SC1329-313 group, as
implied by the detailed analysis that follows. Coordinates correspond to
epoch J2000.0.
\label{f:imh}}
\end{figure*}

\section{Data and Analysis}

XMM-Newton (Jansen et al. 2001) has observed A3562 as a part of the GTO
program of MPE. Tab.\ref{t:ol} details the mosaic of the cluster, where
column (1) is the name of the proposed field, (2) is the assigned XMM
archival name, (3) R.A. and Decl. of the pointing, (4) net Epic-pn exposure
after removal of flaring episodes, (5) pn filter used, (6) XMM-Newton
revolution number. All Epic-pn observations were performed using the
extended full frame mode with a frame integration time of 199 ms.

The initial steps of data reduction are similar to the procedure tested on
other XMM-Newton mosaics and is described in detail in Briel et
al. (2003). Details of our light curve screening are found in Zhang et
al. (2004). The analysis consists of two parts: estimating the surface
brightness and temperature structure of the cluster and verifying it through
spectral analysis. The first part consists in producing temperature
estimates, based on calibrated wavelet-prefiltered hardness ratio maps and
producing the projected pressure and entropy maps. Wavelet reconstruction
(filtering) is used to find the structure and control its significance. The
background is considered differently in imaging and spectral analysis. In
the first cases we use the in-field estimate of the background for every
instrument and pointing, using events furthest from both the optical axis of
the telescope and bright emission zones of the A3562, where instrumental
background should dominate. To subtract the estimated background from the
image we assume no vignetting in spatial distribution of the background,
suitable for the soft proton component (Lumb et al. 2002). Discussion of the
systematics of this method is presented in Henry et al. (2004).

\begin{center}
\renewcommand{\arraystretch}{1.1}\renewcommand{\tabcolsep}{0.12cm}
\tabcaption{\footnotesize
\centerline{XMM Epic-pn log of A3562 cluster observation.}
\label{t:ol}}

\begin{tabular}{lccccc}
\hline
\hline
      &  Obs.       & Pointing  & net    &  pn & XMM \\
      &   ID        & R.A Decl. & exp.    &  Filter        & Orbit\\
      &             & (Eq.2000) & ksec    &         & \\
\hline
f1    & 0105261301  & 203.2725 -31.6917 & 33.5 & Thin   & 567 \\
f2    & 0105261401  & 202.9625 -31.8000 & 7.3  & Medium & 482\\
f3    & 0105261501  & 203.5000 -31.5069 & 7.5  & Medium & 484 \\
f4    & 0105261601  & 203.1250 -31.6236 & 15.3 & Medium & 485\\
f5    & 0105261701  & 203.2083 -31.8694 & 14.1 & Medium & 571 \\
f6    & 0105261801  & 203.5833 -31.7403 & 4.2  & Medium & 574 \\
\hline
\end{tabular}
\end{center}

\includegraphics[width=8.4cm]{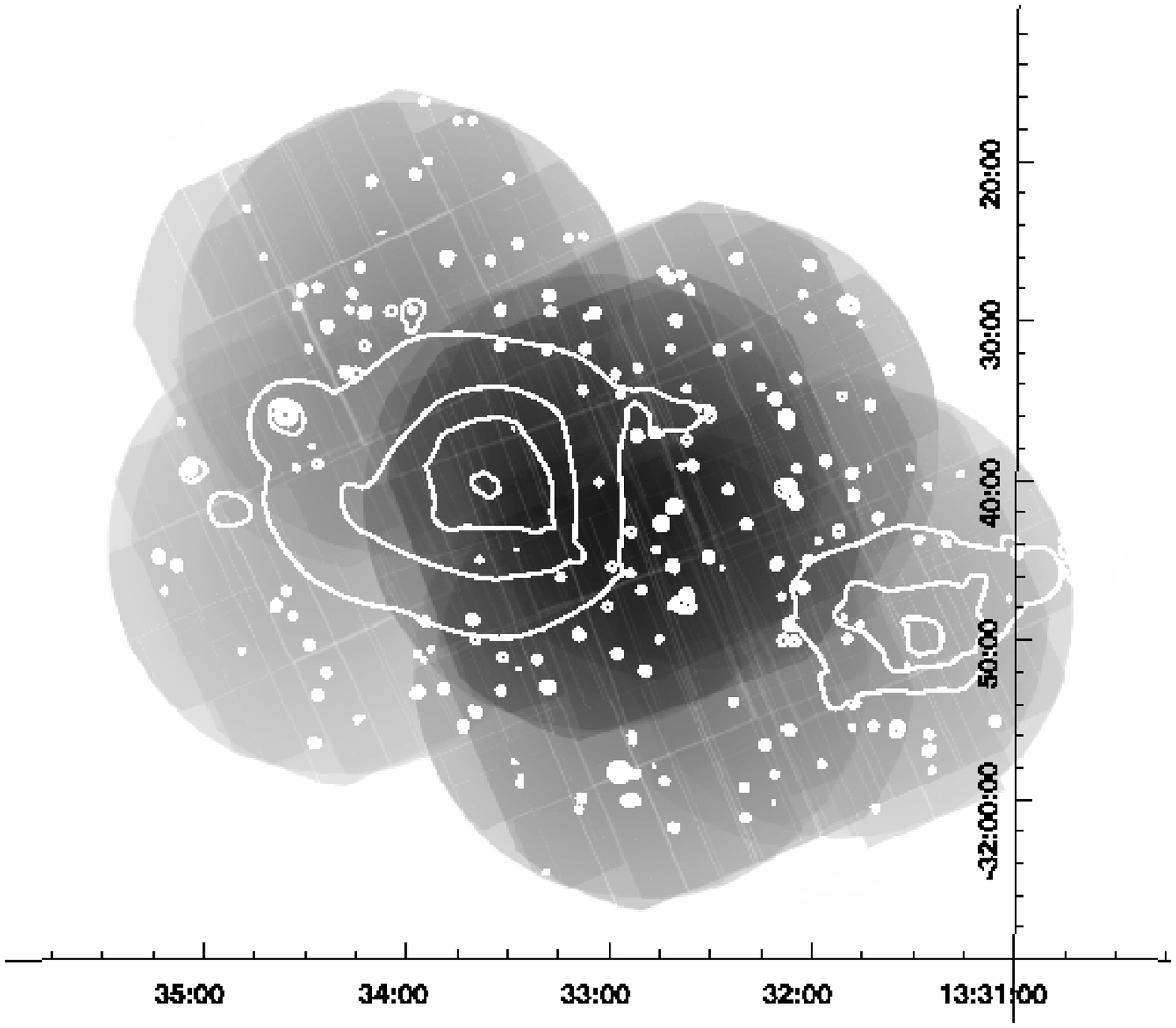}\\
\includegraphics[width=8.4cm]{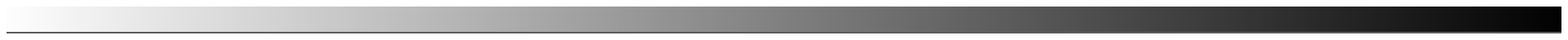}

\figcaption{EPIC (PN+MOS) exposure map of the XMM-Newton survey of A3562
region of Shapley Supercluster, overlayed with contours of equal surface
brightness in the 0.8--2 keV energy band. Coordinates correspond to
epoch J2000.0.
\label{f:exp}}

In Fig.\ref{f:imh} we show a wavelet reconstruction of the XMM mosaic with
contours indicating the emission detected on the smallest scales from
$4^{\prime\prime}$ to $1^{\prime}$. The surface brightness of both A3562 and
the SC1329-313 group reveals distortions on the arcminute (50 kpc) scale,
which were also observed previously by ASCA and BeppoSAX (Akimoto et
al. 2003, Bardelli et al. 2002). The core of A3562 is characterized by a
much shallower gradient in the surface brightness toward the north, as
compared to that to the south and a 100 kpc enhancement in the emission
toward the south-west. The morphology of the group core appears disrupted
and does not reveal a single center  (as indicated by presence of two
peaks in contours displayed in Fig.\ref{f:imh} as well as a zone of high
emission bridging the two peaks, as seen in the color), and on the 100 kpc
scale is characterized by the enhancement in the emission toward the
north-east. As both enhancements in A3562 and in the SC1329-313 group point
toward each other, it appears likely that there exists a connection in their
origins, which we will pursue after a more detailed investigation of the
nature of surface brightness distortions.

The second, spectroscopic part of the analysis uses a mask file, created
based on the results of the hardness ratio analysis described above. The
first application of this technique is presented in Finoguenov et
al. (2004a). Specifics of the A3562 mosaic consists in having observations
performed with a different filter. In our spectral analysis we combine the
events obtained with both filters, use the response matrix without the
filter information, and put the filter information in the arf files. Since
in the merged event file, the filter information will be overwritten by one
of the observations, we calculate the arf for merged observations with the
medium filter (using a XMMSAS task clarf, developed by A. Finoguenov). The
arf is calculated for the observation with the thin filter separately and
later added. The weighting is achieved within clarf through a comparison of
the exposure time stored in the PHA file and calculated for the input event
file. For background removal we use the background accumulation, obtained
with the medium filter, since subtle differences with the accumulation with
the thin filter due to galactic foreground emission are not important for
the analysis of the core of A3562, where the thin filter was used. We also
use the 0.8--8 keV band in the analysis to avoid the possible complications
due to a soft excess, which further reduces the differences in the
backgrounds between the observations performed with the thin and medium
filters.

\section{Results}

\begin{figure*}

\includegraphics[height=19.cm,angle=-90]{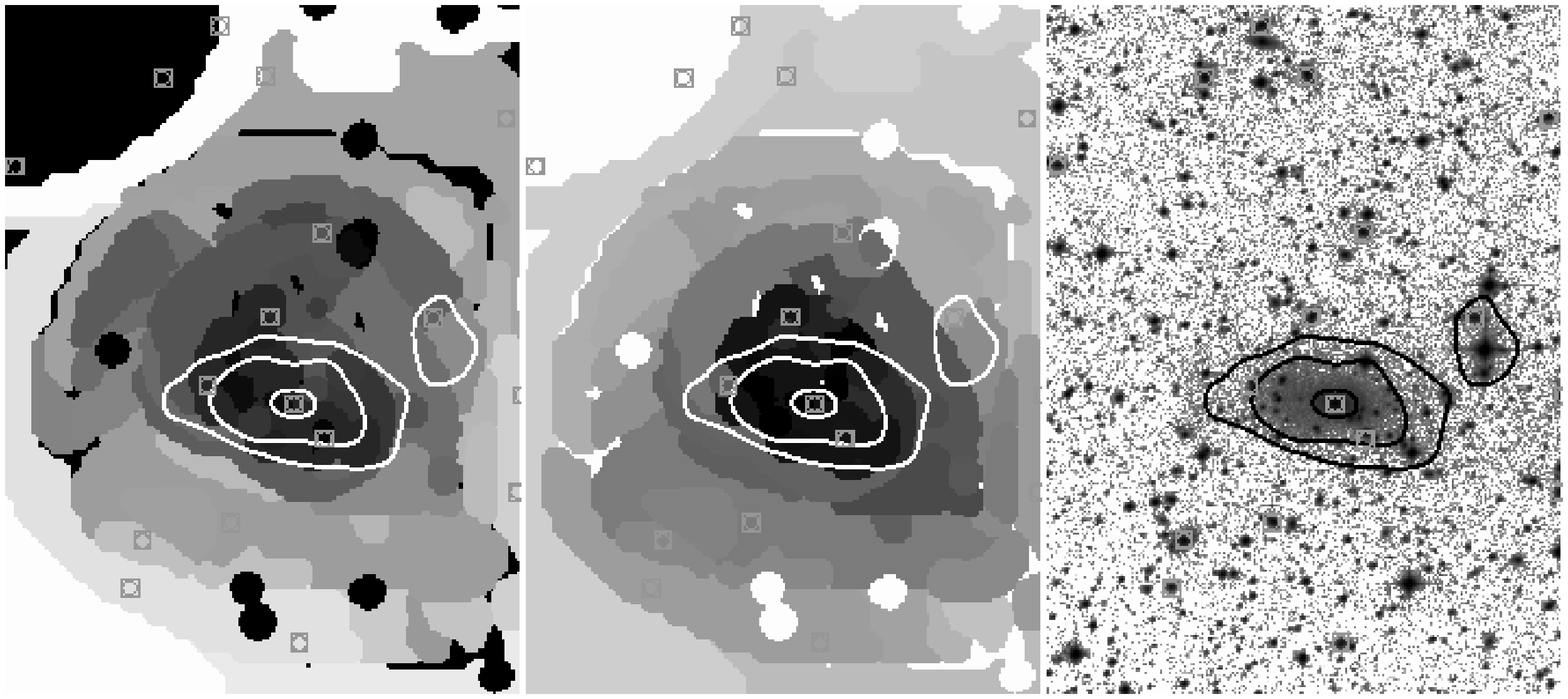}\\
\includegraphics[width=19.cm]{plots/grey_cbar.ps}

\figcaption{From left to right, projected entropy, pressure and DSS2 R band
image of A3562 all overlaid with the contours of equal diffuse optical light
intensity. The left panel also displays the position of the cluster galaxies
with measured redshifts.
\label{f:dss-all}}

\end{figure*}

The advantage of XMM-Newton over previous missions consists in the ability
to provide detailed temperature maps, which allows a study of pressure and
entropy state of the cluster gas. In Finoguenov et al. (2004a) and Briel et
al. (2003) we have developed a simple technique on how to select the regions
for the spectral analysis, using both surface brightness and hardness ratio
as an input. Given the observational setup, which is further
  illustrated in Fig.\ref{f:exp}, there is a limited ability to
study the structure of the SC1329-313 group, as the total number of counts
there provides only three spectra. The deep exposure on A3562 allows us to
produce almost a hundred of spatially independent temperature estimates. So,
while we list the properties of both A3562 and the SC1329-313 group in
Tab.\ref{t:xray} later in this section, we will concentrate on A3562 in the
discussion of the 2d structure.

In our spectral analysis we use the APEC plasma code to measure the
temperature, element abundance (assuming the solar abundance ratio of Anders
\& Grevesse 1989) and emission measure. By making an estimate of the volume,
occupied by each emission component, it is then possible to recover the
pressure and the entropy in absolute units. In general, volume estimates are
not immediately taken as convincing, although errors resulting from it are
small, typically on the 20\% level (for details see Finoguenov et al. 2004a;
2004b; Henry et al. 2004). We also study projected pressure and entropy maps,
where instead of the density we use the square root of the emission measure,
which then includes the projected length. The second order effect due to
multi-temperature components seen in projection was found to reduce the
pressure enhancements due to shocks, as hotter components have lower
emission measures. Also, due to the weighting of the temperature components
by the density squared, X-ray analysis is mostly sensitive to the state of
the most densest part in the projected length.

With the limitations given above, in Fig.\ref{f:dss-all} we display the
projected pressure and entropy maps derived in the spectral analysis. The
peak of pressure and entropy is centered on the galaxy identified with the
BCG of A3562 cluster by Postman \& Lauer (1995), which has a
heliocentric velocity of $14653\pm15$ km/s.

Compared to the distribution of stars in the bright cluster galaxy (BCG)
envelope, shown in overlay, both pressure and entropy of the gas is extended
towards the north-east. The entropy map has an even larger extension to the
north-east compared to the pressure map and breaks into two distinct tails
there. The pressure map has an enhancement west to the galaxy. Thus, the
core of A3562 appears to be disturbed in a number of ways. Given the
importance of the large-scale entropy distribution for the picture of A3562,
in Fig.\ref{f:smap} we present the pseudo entropy map, derived as $T /
\sqrt[3]{I}$ (Churazov et al. 2003; Briel et al. 2003) where $I$ is a
surface brightness in the 0.8--2 keV and $T$ is derived from the calibrated
hardness ratio map. This map illustrates the observed shape of the low
entropy tails in A3562.

\includegraphics[width=8.cm]{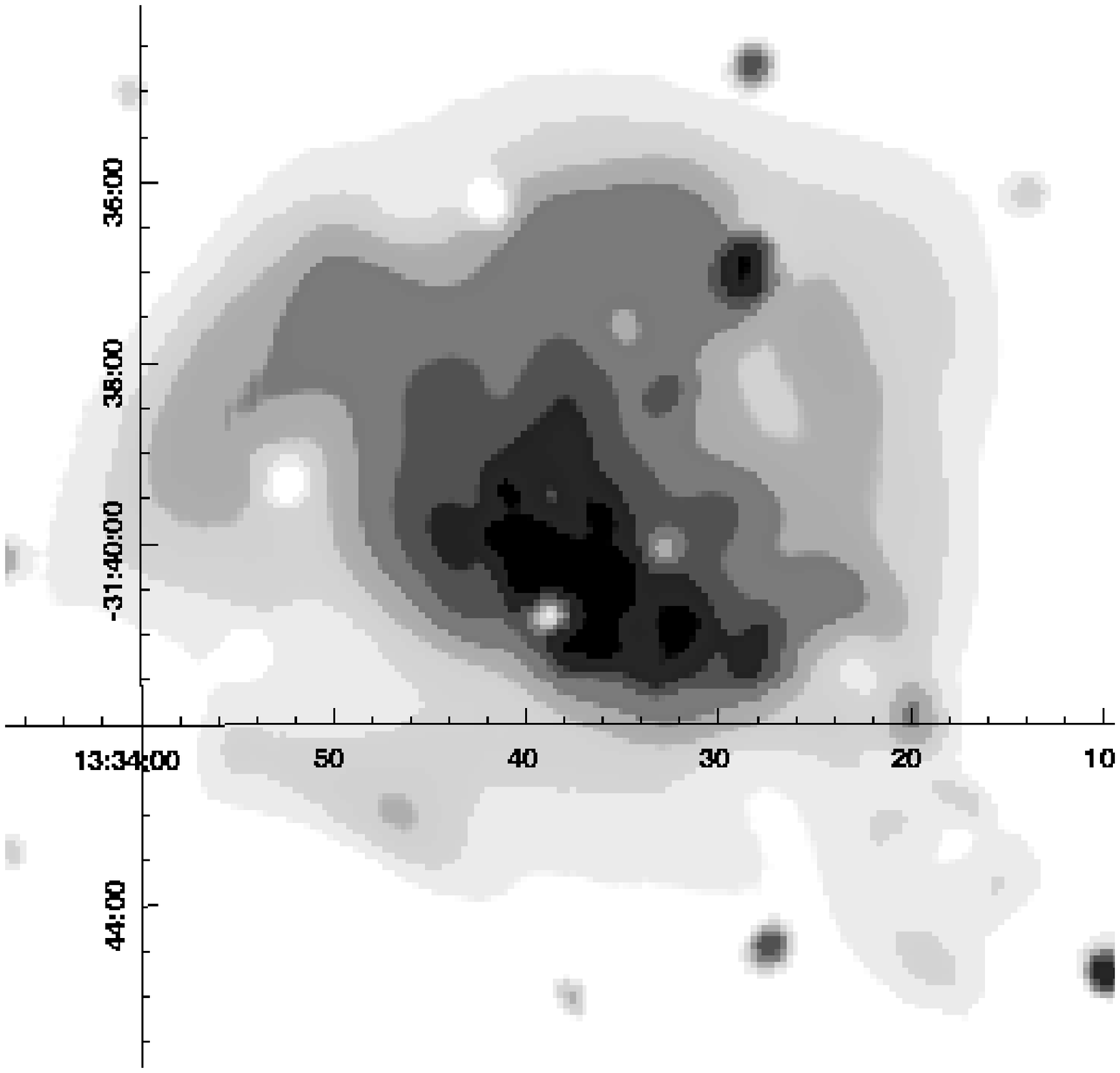} \\
\includegraphics[width=8.cm]{plots/step_grey_cbar.ps}
\figcaption{Entropy map of A3562, derived as $T / \sqrt[3]{I}$. Coordinate
grids are shown for the J2000.
\label{f:smap}}

\begin{table*}
\begin{center}
\renewcommand{\arraystretch}{1.1}\renewcommand{\tabcolsep}{0.12cm}
\tabcaption{\footnotesize
\centerline{X-ray mass averaged properties of A3562.}
\label{t:xray}}

\begin{tabular}{cccccc}
\hline
\hline
 Radii & kT  & Z        & S        &P       & Gas mass \\
 kpc   &  keV  & $Z_\odot$& keV      & $10^{-12}$   & $10^{11}$\\
        &      &      & cm$^{2}$&ergs cm$^{-3}$& $M_\odot$ \\
\hline
\multicolumn{4}{c}{A3562}\\
0--130  & $4.20\pm0.04$&$0.46\pm0.02$&$135\pm2$&$ 58.44\pm0.74$&$8.29\pm0.06$\\
130--300& $4.33\pm0.07$&$0.35\pm0.03$&$230\pm5$&$27.68\pm0.46$&$14.68\pm0.14$\\
0--300  & $4.28\pm0.04$&$0.39\pm0.02$&$195\pm3$&$38.78\pm0.40$&$22.97\pm0.15$\\
300--600&$3.65\pm0.08$&$0.18\pm0.03$&$584\pm20$&$6.09\pm0.11$&$136.44\pm2.13$\\
0--600 &$3.74\pm0.07$&$0.21\pm0.02$&$528\pm18$&$10.80\pm0.11$&$159.41\pm2.14$\\
\multicolumn{4}{c}{SC1329-313}\\
0--130  & $3.00\pm0.19$&$0.28\pm0.10$&$202\pm19$&$13.06\pm1.02$&$5.39\pm0.25$\\
130--600&$3.13\pm0.10$&$0.09\pm0.03$&$659\pm25$&$2.97\pm0.12$&$151.88\pm2.16$\\
0--600  &$3.12\pm0.09$&$0.09\pm0.03$&$643\pm24$&$3.31\pm0.12$&$157.27\pm2.18$\\
\hline
\end{tabular}
\end{center}
\end{table*}

Before proceeding with the interpretation, we first point out a degeneracy
in the available solutions. Enhancements in the pressure maps could be
treated either as dark matter distribution or hydrodynamic effects. The
pressure enhancements discussed here occur at the 10--40\% level. While
these could indicate the presence of a weak shock, it is also feasible to
obtain such pressure enhancements within subsonic gas motions, as soon
as there is a change in the velocity of the gas. These processes are
described by the Euler's equation (e.g. Landau \& Lifshitz 1959).
\begin{equation}\label{eq:euler}
{\partial {\bf v} \over \partial t} + ({\bf v \cdot \nabla}) {\bf v} =
-{{\bf \nabla} p \over \rho} - {\bf \nabla \Phi}
\end{equation}
Here we neglect the action of viscosity on diminishing the velocity
gradients as well as contributions of the magnetic field, which otherwise
enter as
\[ \nu \nabla^2 {\bf v} \] 
and
\[ -{{\bf \nabla} B^2 \over 8 \pi \rho } + {({\bf B \cdot \nabla}){\bf B}
  \over 4 \pi \rho} \] to the right hand side of the Eq. 1, respectively.
  We denote ${\bf B}$ to be the magnetic field, ${\bf v}$ as the velocity,
  $p$ as the thermal pressure, $\rho$ as the density, $\nu$ as the
  viscosity, ${\bf \Phi}$ as the gravitational potential.

Small changes in the entropy of the gas, characteristic of weak shocks, does
not allow a robust differentiation between the two
possibilities. Essentially, Eq.\ref{eq:euler} illustrates that the gas
pressure traces the generalized stress tensor.

In the case of a bow-shock, there are further geometrical constrains on the
appearance of the flying body, such as the angle of the bow-shock, which are
not observed. The disturbed morphology of the SC1329-313 group south-west to
A3562 as well as the presence of compressed gas to the south from A3562
argues in favor of a recent passage of the group.  The X-rays of SC1329-313
group are centered on AM 1328-313, which is a galaxy with velocity 12928 km/s
(NED). However, the galaxy has two distinct cores and redshift measurements
reported with slight offset have higher velocity by 300 km/s.

The motion of the cluster core is likely to represent an oscillation in
response to the passage of the subcluster, observed west to the A3562. On
even large scale there is a good degree of symmetry in the pressure relative
to the position of the BCG, yet with 30\% pressure excess to the south. The
entropy map on the large scale is also elongated in the north-south
direction. One possible explanation of the observed picture would be that
the cluster exhibits oscillations on two scales, on the larger scale (0.1
degree) the oscillation occurs in the north-south direction. On the smaller
scale there is a significant east-west asymmetry in the pressure map.  The
length of the pressure enhancements to the east, where it enters into zones
of lower surface brightness, supports an association of this pressure
enhancement with the presence of a lateral shock caused by the passage of
the SC1329-313 group.

Typical pressure enhancements are 40\% and, if associated with the sloshing,
imply that the oscillation velocity is 15\% of the sound speed. The
distribution of the entropy follows the oscillations yet at a fraction of
the sound speed, typically 10--20\% (e.g. Ricker \& Sarazin 2001). Since the
lowest entropy is found almost at the location of BCG, this scenario seems
to be very likely. So, our conclusion is that the pressure enhancements are
associated with a deceleration associated with the oscillation of both the
BCG and the cluster. The BCG has stopped going north-east, the cluster has
almost reached the most southern point in the oscillation period.  We
estimate the oscillation frequency of the BCG to be $\sim 1$ Gyr$^{-1}$.

A question of the origin of the low entropy tails within the assumed
subsonic motion of the A3562 core puts a tail of A3562 in correspondence
with the remnant of the action of the Rayleigh-Taylor instability on
disruption of cores of the cluster. In this scenario the change in the core
velocity at the oscillation apogee produces a pull that balances the
gravitation force and leads to an escape of low entropy gas. In this picture
A3562 is a more advanced stage compared to A3667, where such disruption is
also suggested (Briel et al. 2003). The duration of time for which such a
feature is observed is determined by convective settlement of the gas to the
potential minimum, as discussed elsewhere in this paper. One can not
exclude, however, that the process is ongoing, which requires a different
frequency (or a phase difference) for the oscillation occurring with the
increasing distance from the cluster center, i.e. when the core flies to the
south, the bulk of the cluster flies to the north, as has been suggested
above already. Discussion of the galaxy substructure in A3562,
presented in \S\ref{s:merger} is a strong support for this scenario. Yet
another possibility is proposed by Heinz et al. (2003), who specifically
considered the case of A3667. Within their scenario, the pull resulting in
the Rayleigh-Taylor instability is created by the passage of the shock,
which in the case of A3562 should be associated with interaction of the
SC1329-313 group with the outskirts of A3562. In a summary, while it is
rather certain that the Rayleigh-Taylor instability is the key process that
explain the low entropy tails observed in A3562, there are a number of
processes that could lead to such an observed effect.

Tab.\ref{t:xray} lists the mass averaged properties of both the A3562
cluster and the SC1329-313 group. Col (1) specifies the radius, (2)
Temperature, (3) element abundance, (4) entropy, (5) pressure, (6) the gas
mass involved in weighting. As no account for the projected components is
made, this is not the total gas mass within the radii quoted.

Our data on the SC1329-313 group show an interestingly high entropy level,
indicating that groups in superclusters are systematically underluminous, in
agreement with a study of Hanami et al. (1999), based on the $L_x-T$
relation. We attribute it to a high entropy of the intracluster gas in
superclusters, possibly connected to the structure formation shocks.

\subsection{Key properties of A3562}

\begin{table*}
{
\begin{center}
\footnotesize
{\renewcommand{\arraystretch}{0.9}\renewcommand{\tabcolsep}{0.09cm}
\caption{\footnotesize
\centerline{Properties of main regions of A3562.}
\label{t:main}}
\begin{tabular}{ccccccccccccc}
 \hline
 \hline
N & Name &$kT$ &$Fe/Fe_\odot$& $\rho_e$ & S & P, $10^{-12}$ & $M_{\rm gas}$ & $r_{\rm min}$ & $r_{\rm max}$\\
  &  & keV  &   &  $10^{-4}$ cm$^{-3}$& keV cm$^2$ & ergs cm$^{-3}$ &
  $10^{12} M_\odot$ & kpc & kpc  \\
\hline
 2& pressure core   &$ 4.2\pm0.1$&$0.53\pm0.04$&$80.0\pm1.0$&$105\pm 2$&$53.6\pm1.1$&$ 0.1\pm0.0$&  0&  65\\
 3& entropy core    &$ 4.1\pm0.1$&$0.41\pm0.03$&$43.5\pm0.4$&$154\pm 3$&$28.7\pm0.5$&$ 0.4\pm0.0$& 30& 131\\
 4& main-1          &$ 4.6\pm0.1$&$0.39\pm0.04$&$36.3\pm0.4$&$195\pm 5$&$26.8\pm0.7$&$ 0.3\pm0.0$& 38& 158\\
 5& dove tail-1     &$ 4.0\pm0.1$&$0.44\pm0.04$&$19.7\pm0.3$&$253\pm 6$&$12.5\pm0.3$&$ 0.6\pm0.0$& 92& 265\\
 6& dove tail-2     &$ 3.8\pm0.1$&$0.28\pm0.05$&$ 7.8\pm0.1$&$452\pm15$&$ 4.7\pm0.2$&$ 1.9\pm0.0$&201& 628\\
 7& southern shock-2&$ 5.1\pm0.2$&$0.28\pm0.08$&$10.8\pm0.3$&$489\pm25$&$ 8.9\pm0.5$&$ 0.5\pm0.0$&223& 377\\
 8& pressure ridge  &$ 4.7\pm0.1$&$0.32\pm0.04$&$13.8\pm0.2$&$376\pm11$&$10.3\pm0.3$&$ 0.6\pm0.0$&117& 419\\
 9& main-2          &$ 4.6\pm0.1$&$0.24\pm0.03$&$13.2\pm0.1$&$383\pm10$&$ 9.7\pm0.3$&$ 0.9\pm0.0$&101& 363\\
10& southern shock-1&$ 4.8\pm0.1$&$0.27\pm0.04$&$15.0\pm0.2$&$369\pm10$&$11.6\pm0.3$&$ 1.0\pm0.0$& 97& 322\\
11& main-3          &$ 4.2\pm0.1$&$0.26\pm0.03$&$ 4.5\pm0.0$&$717\pm13$&$ 3.1\pm0.1$&$10.2\pm0.1$&274& 713\\
12& outskirts-1     &$ 2.1\pm0.1$&$0.13\pm0.05$&$ 2.9\pm0.1$&$484\pm29$&$ 1.0\pm0.1$&$ 6.1\pm0.2$&675& 967\\
13& outskirts-2     &$ 1.9\pm0.2$&$0.12\pm0.06$&$ 1.7\pm0.1$&$612\pm65$&$ 0.5\pm0.1$&$13.6\pm1.1$&957&1420\\
\hline
\end{tabular}
}
\end{center}
}
\end{table*}

\begin{figure*}

\includegraphics[width=19.cm]{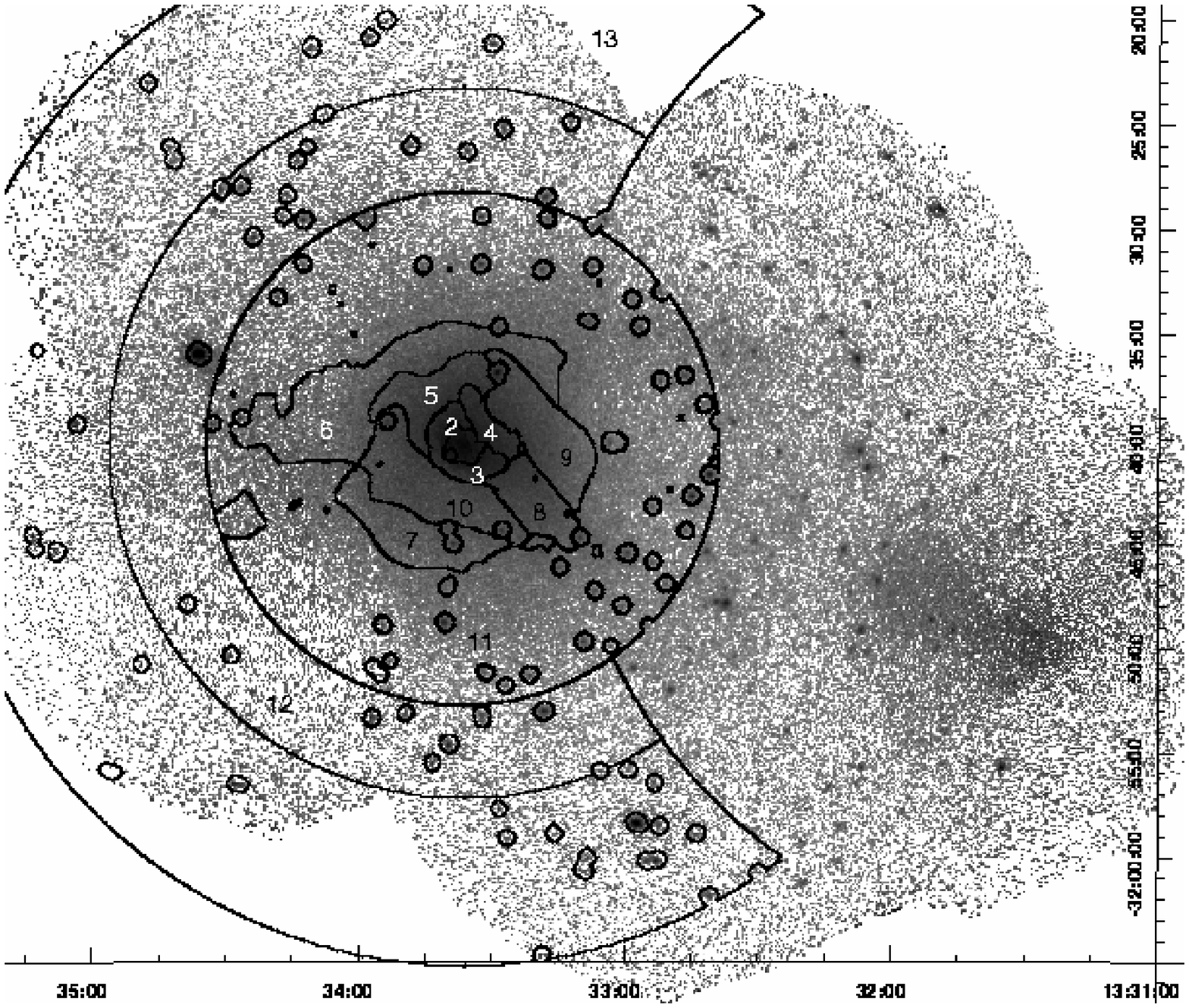}\\
\includegraphics[width=19.cm]{plots/grey_cbar.ps}

\figcaption{Background subtracted, exposure corrected image of A3562 in the
  0.8--2 keV band. Contours show the location of the regions selected for
  the spectral analysis, reported in Tab.\ref{t:main}.
\label{f:main}}

\end{figure*}

In order to tabulate the basic properties of the cluster we extracted the
spectra from large zones that according to both the hardness ratio and the
fine spectral analysis have similar pressure and entropy. We used the 1--10
keV energy band for the spectral analysis, as compared to 0.75--10 keV band
used for the rest of the paper to avoid bias to lower temperature components
resulting in this crude region selection. We list the properties obtained
this way in Tab.\ref{t:main} with their $\pm1\sigma$ errors for one
parameter of interest. Col. (1) labels the region according to
Fig.\ref{f:main}, with the name of the region reported in (2), (3) lists
temperature in keV, (4) iron abundance in solar photospheric units of Anders
\& Grevesse (1989). Derived quantities, that assume a center of A3562 at
203.40080, -31.67145 (equinox J2000.) and use an estimation of the projected
length, as described above are reported in cols.(5--8). These are electron
density, entropy, pressure and the (local) gas mass. Cols. 9--10 report the
minimal and maximal distance to the extraction area. No account for the gas
mass not associated with the directly observed component was
attempted. Region 1, which is embedded in the region 2, is not reported in
the Table as it turned to be an AGN from the X-ray spectral properties.

In Fig.\ref{f:s}--\ref{f:p} we compare the obtained entropies and pressures
with known scaling relations. In the adopted cosmology, estimate of the
$r_{500}$ is 0.91 Mpc, where the scaling of $r_{500}$ with temperature is
taken from Finoguenov et al. (2001) and the temperature of A3562 is taken as
4.3 keV. As both entropy and $r_{500}$ are scaled by a similar power of
temperature, the deviation of data points in Fig.\ref{f:s} from the scaled
entropy profile of Ponman et al. (2003) is not very sensitive to the adopted
temperature. The normalization of the scaled entropy is taken to correspond
to the clusters with temperature below 5 keV in Ponman et al. (2003).

The data show agreement with the scaling relations. The core of the cluster
has higher entropy with respect to the scaling, which is typical also for
the relaxed clusters (e.g. Pratt \& Arnaud 2003). The dove-tale-like
structure has lower entropy, but at the same time a typical pressure
expected for its distance to the cluster center. The region (5) has, in
addition, an iron abundance, which is similar to the core of the cluster,
but higher than the rest. Comparison between regions (6) and (7) reveals
almost a factor of two higher pressure of the latter, compared to lower
(though comparable) entropy of the former. This results from pressure
enhancement to the south, but mostly from extension of the low-entropy tails
into the zones of low pressure to the north from the cluster
center. Pressure discontinuities correspond to a transition zone between
cluster core and the rest of the cluster, so it is difficult to estimate the
unperturbed pressure value. However, comparing zones 8--10, one can see that
they have a similar distance to the center and similar entropy. The ridge in
the image, that corresponds to region (8), shows however only a marginal
pressure enhancement, while the pressure enhancement to the south is clear
and correspond to a 20\% pressure enhancement. It is likely that spatial
resolution of XMM-Newton data prevents us to perform a clean separation
required to detect the hot component associated with the ridge, seen with
Chandra (Ettori, S. 2004, private communication).

In order to provide some insights on the origin of the large-scale emission
detected by Kull \& B\"ohringer (1999), we have constructed two regions,
(12--13), that reach $25^\prime$ distance from the A3562 center. Background
subtraction becomes an issue for these regions and we have tried several
background accumulations to test the robustness of our results. In addition
to the level of instrumental background, as revealed by the data in the
5--12 keV band, we find a need to include an additional 0.2 keV component
with solar element abundance typical to galactic halo emission. The XSPEC
normalization, K, of this component per square arcminute is $3.4\times
10^{-6}$. In Tab.\ref{t:main} we list the hot thermal component, also seen
in the spectral analysis. Given the complexity of the background situation
for XMM-Newton, usual approach is to look for supporting measurement with
satellites having lower background. We point out that the reported BeppoSAX
LECS+MECS results of Ettori et al. (2000) yield similar temperature and
pressure for parts of the A3562 cluster corresponding to our regions 12 and
13, therefore providing support that the component reported here originates
from A3562. This component, having temperature of 2 keV is colder compared
to the 3 keV temperature observed on the western part of A3562 (see
Tab.\ref{t:xray}). State of the gas in regions 12--13 is characteristic of
virial regions of clusters with already maintained pressure balance, but
lower entropy level indicating either a survival of low-entropy gas or an
incomplete thermalization. We point out that this situation has been
observed in simulations before (e.g. Evrard et al. 1996) and correspond to
the accretion zone of the cluster between virial radius and $r_{500}$. While
such entropy behavior is not unique to A3562, (e.g. Finoguenov et al. in
preparation) we point out that deviations in the pressure profile from the
-2.5 slope power law radial behavior are only seen in A3562 and might be a
specific of supercluster environment.

\includegraphics[width=8cm]{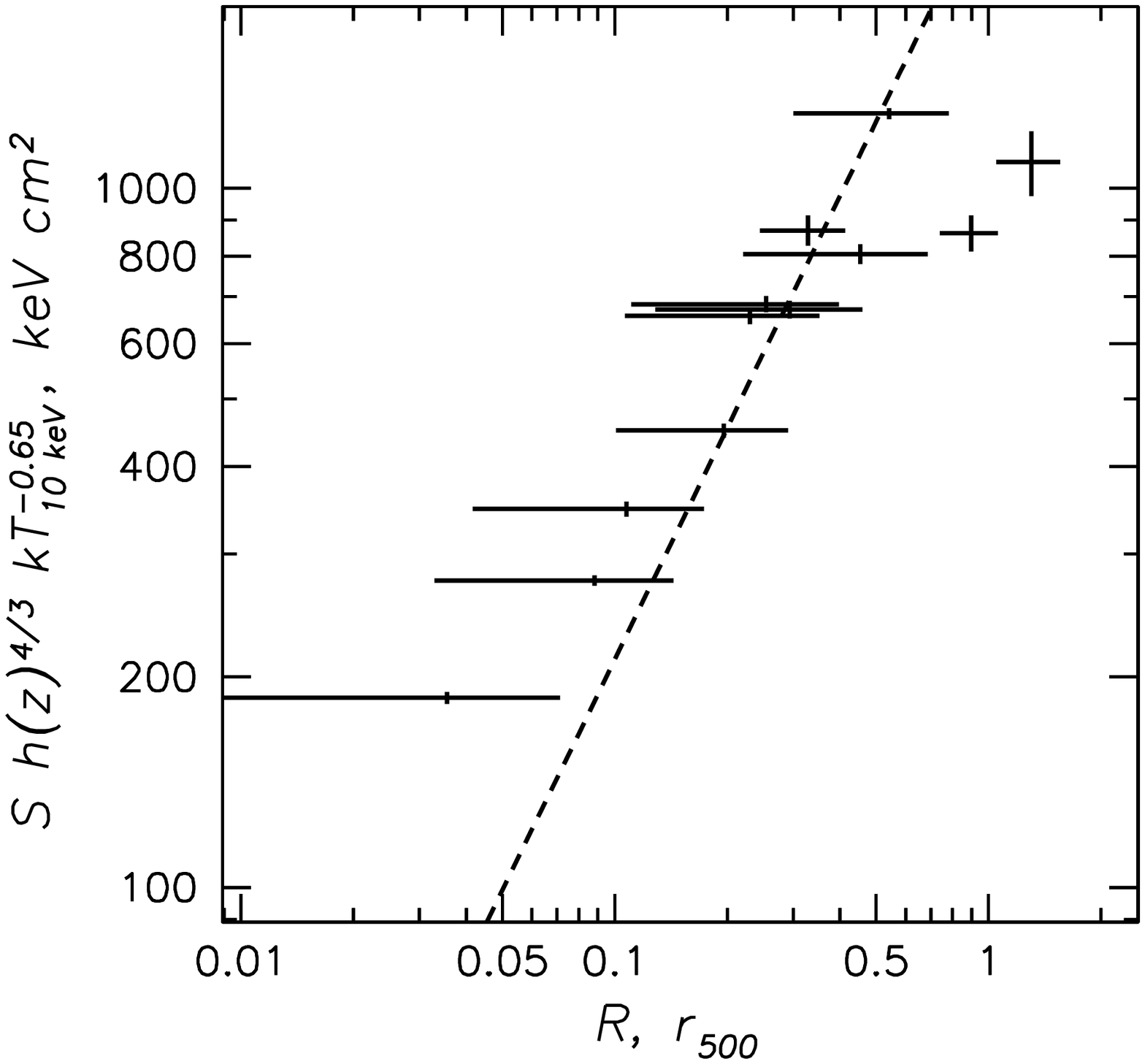}
\figcaption{Entropy profile of the main zones in A3562. Universal
entropy scaling relation of Ponman et al. (2003) is shown with a dashed
line.
\label{f:s}}

\includegraphics[width=8cm]{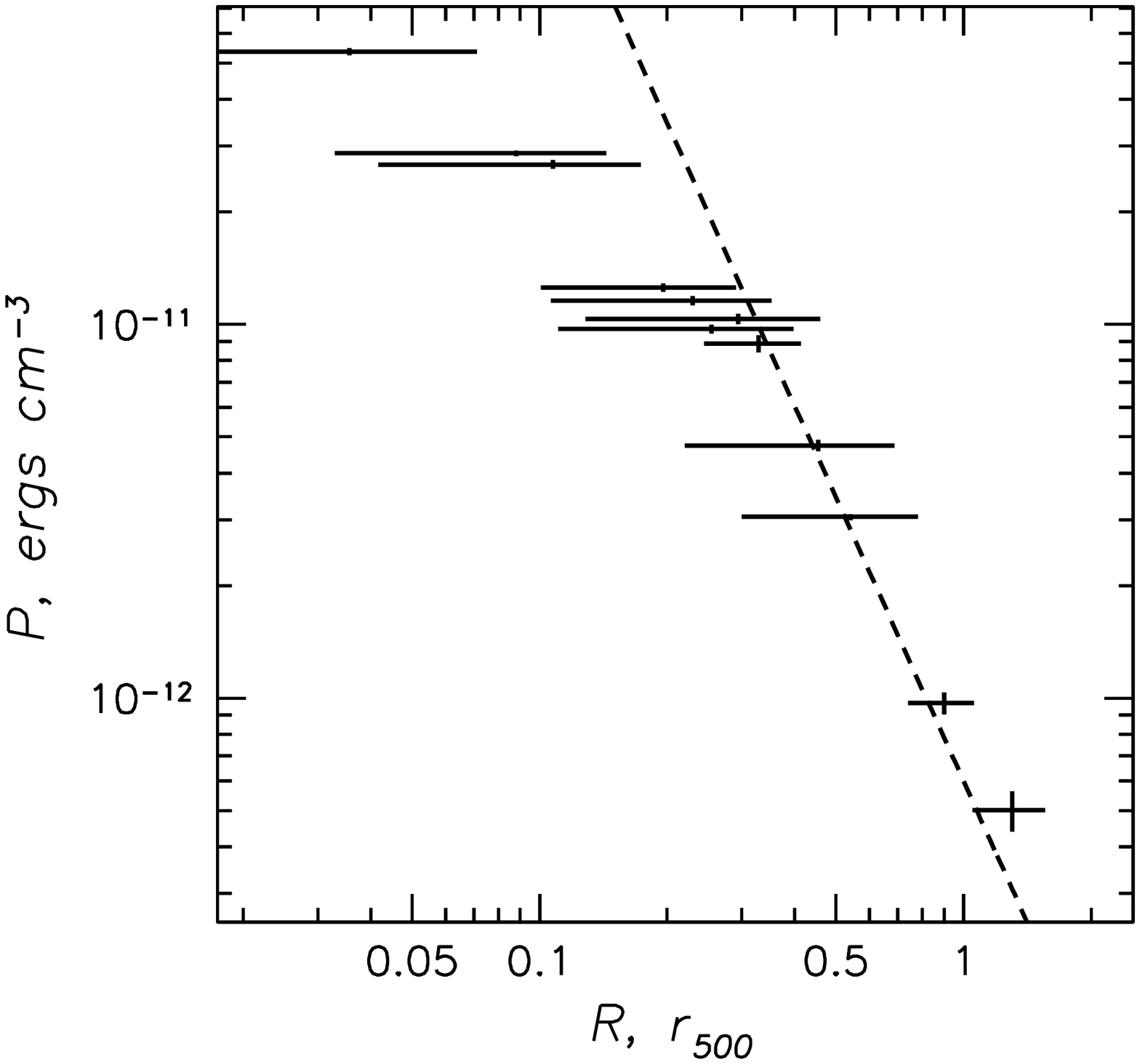} \figcaption{Pressure profile of
the main zones in A3562. A pressure slope of -2.5, typical for the relaxed
clusters is shown with a dashed line. Normalization is done using the
outmost point.
\label{f:p}}

\subsection{Geometry of the merger}\label{s:merger}

The observed sloshing of gas, as a result of oscillations triggered by the
passage of the SC1329-313 group, suggests that the group was {\it initially}
flying in the western direction. After passage at perigee with A3562, the
SC1329-313 group has been deflected acquiring a velocity component towards
the A3562. A similar momentum in the opposite direction has been acquired by
the cluster core, as for the rest of the cluster the fluctuations in the
gravitational potential are negligible. Oscillations of the cluster core
will be counterbalanced on the larger scale by the cluster to conserve the
momentum initially acquired. An interesting confirmation of this statement
comes from a comparison to a study of Bardelli et al. (1998), indicating
that the major 3d galaxy concentration in A3562, T561, is displaced north to
the BCG, and coincides with position of the dove-tail.  The recession
velocity of T561 is $14527^{+168}_{-229}$, similar to the velocity of the
BCG, supporting therefore a subsonic motion of the later.

We suggest that the SC1329-313 group has passed north to A3562. Then A3562
is just finishing the first oscillation period. An extra half a period is
required in the scenario when the group passes the cluster on its southern
side. Lack of low-entropy tails in the south argue against it. Also the
image of A3562 appears to be more disturbed in the north, which we relate to
shock heating due to interaction of gaseous halos. In addition, the
structure of the group suggests that it has a component in the southern
direction.

It appears interesting to compare the conditions of the gas on the
large-scale north and south from the cluster to see if it fits into the
above suggested picture. The observed separation of the SC1329-313 group and
A3562 is a factor 7--10 larger than the oscillation amplitude, suggesting
that the characteristic Mach number of the interaction was in the 1.1--1.5
range (velocity of the group in the scale of the sound speed of the A3562
gas).

An apparent need for stronger interaction to explain the shock compared to
the projected separation between A3562 and the SC1329-313 group, could
partly be resolved in the significant projected velocity component, which is
$1200$km/s when we formally take the velocity difference between the major
galaxies. A large projected velocity difference also explains why the
pressure feature appears only to the west from A3562.

The high velocity of the SC1329-313 group could further be checked by the
comparison between the expected disruption of the group, occurring at the
pressure level of $P_{ram}=7\pm1 \times 10^{-12}$ ergs cm$^{-3}$ and the
suggested supersonic motion through a medium of $\rho_{\rm ICM}=2.4\pm0.2
\times 10^{-4}$ cm$^{-3}$ electron density ($P_{ram}=\rho_{\rm ICM} v_{\rm
group}^2$). The velocity of $1700\pm150$ km/s, implied by this scenario, is
in reasonable agreement with the other arguments.

\subsection{Comparison to radio}

It has been found for A754 that the faint radio emission correlates with the
pressure map (Henry et al. 2004). In Fig.\ref{f:nkt-r} we provide a
comparison of the radio emission (NVSS, 20 cm) with the pressure map of
A3562. The point-like source represents the radio galaxy traveling through
the core of A3562. There is again an association of the radio emission on
the $\mu$Jansky level with the pressure peak on the pressure level exceeding
the $5\times 10^{-11}$ ergs cm$^{-3}$. Importance of the A3562 radio halo is
that it is one of the weakest halos found, while A3562 is one of the coldest
clusters with the presence of radio halo. The spectral index of the halo
reveals its relatively young age (Venturi et al. 2000, 2003). Among abundant
reacceleration mechanisms, discussed in the literature, a suggestion of
Venturi et al. (2003) for A3562 on reviving of the radio plasma, seeded by
the head-on radio galaxy, seems to be supported by complicated pressure
structure in the core of A3562, as indicated by our data. Thus, the extent
of radio emission in A3562, might be caused by the initial distribution of
the seeds and not by the extent to which the shocks are propagated. Further
discussion of the link between the XMM results and the radio structure of
A3562 will be presented elsewhere.

\includegraphics[width=8.cm]{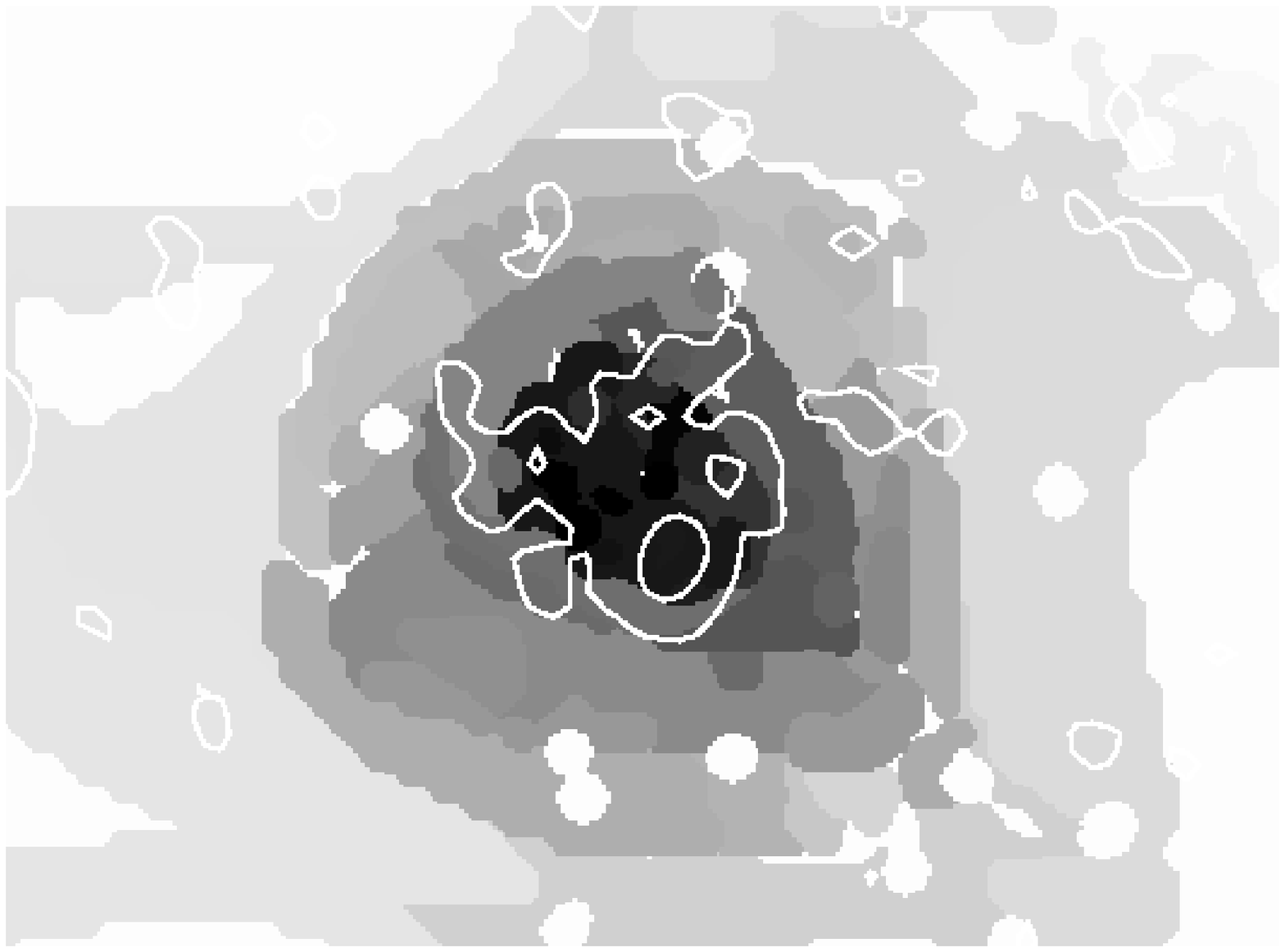}\\
\includegraphics[width=8.cm]{plots/grey_cbar.ps}
\figcaption{Pressure map of A3562, overlaid with the radio contours.
\label{f:nkt-r}}

\section{Summary}

The paper describes the results of XMM-Newton mosaic of A3562 and a nearby
SC1329-313 group. The key measurements discussed are the tail of low entropy
gas to the north of the core of A3562, 20\% pressure enhancement to the
south, and a distorted shape of the nearby SC1329-313 group. The presence of
the low entropy tails is related to the action of Rayleigh-Taylor
instability. The cause of both instability and the pressure enhancement
could either be core oscillations induced by the passage of the group or by
ongoing propagation of the shock wave through the A3562, which is induced by
an interaction of the group with the outskirts of A3562. It is likely that
the observed picture is the result of a combination of the two
scenarios. Consideration of the projected velocity of the SC1329-313 group,
a comparison of distorted shape of the group with inferred ram pressure, all
support the supersonic motion of the group through the outskirts of A3562.

\begin{acknowledgments}
The paper is based on observations obtained with XMM-Newton, an ESA science
mission with instruments and contributions directly funded by ESA Member
States and the USA (NASA). The XMM-Newton project is supported by the
Bundesministerium f\"{u}r Bildung und Forschung/Deutsches Zentrum f\"{u}r
Luft- und Raumfahrt (BMBF/DLR), the Max-Planck-Gesellschaft (MPG) and the
Heidenhain-Stiftung, and also by PPARC, CEA, CNES, and ASI. SRON is
supported financially by NWO, the Netherlands Organization for Scientific
Research. AF acknowledges support from BMBF/DLR under grant 50 OR 0207 and
MPG. AF thanks the Joint Astrophysical Center of the UMBC for the
hospitality during his visit. Authors thank Pat Henry, Stefano Ettori and an
anonymous referee for useful suggestions that significantly improved
presentation of the material.
\end{acknowledgments}

\end{document}